# The Hourglass Revolution: A Theoretical Framework of AI's Impact on Organizational Structures in Developed and Emerging Markets

Krishna Kumar Balaraman[1] & Venkat Ram Reddy Ganuthula[2]

[1]IFMR Graduate School of Business, Krea University; [2]Indian Institute of Technology Jodhpur

**Abstract**

This paper presents a theoretical framework examining how artificial intelligence (AI) transforms organizational structures, introducing an "hourglass" configuration that emerges as AI assumes traditional middle management functions. The analysis identifies three key mechanisms algorithmic coordination, structural fluidity, and hybrid agency that demonstrate how AI enables organizational forms transcending traditional structural boundaries. These mechanisms illustrate how AI enables new modes of organizing to go beyond existing structural boundaries. Drawing on institutional theory and digital transformation research, we examine how these mechanisms operate differently in developed and emerging markets, producing distinct patterns of structural transformation. Our framework offers three important theoretical contributions: (1) conceptualizing algorithmic coordination as a unique form of organizational integration, (2) explaining how structural fluidity allows organizations to achieve stability and adaptability at the same time, and (3) the theoretical argument that hybrid agency surpasses traditional, human centric forms of organizational capabilities. Our analysis shows that while the move to AI enabled strategies overall seems quite global, successful application will need to pay sufficient attention to the technological capabilities, cultural dimensions, and contexts of the market.

## Introduction

The integration of artificial intelligence (AI) fundamentally transforms organizational structure and decision-making processes (Brynjolfsson & McAfee, 2017; Agrawal et al., 2018). While early AI implementations focused primarily on task automation, current capabilities are driving unprecedented changes in organizational design and decision-making architecture (Krakowski et al., 2022). Three key technological advances drive this transformation: exponential growth in computational power, unprecedented data availability, and increasingly sophisticated machine learning algorithms (Kolbjørnsrud et al., 2016). These advances enable a shift from traditional enterprise computing focused on efficiency to ubiquitous computing, characterized by seamless integration of physical and virtual work environments (Cascio & Montealegre, 2016).

An organization's framework is much more than the role systems and processes that are defined; it includes repeated interaction patterns bearing on the working of the organization (Ahmady et al., 2016). Therefore, the rise of artificial intelligence brings forth some very fundamental questions about such frameworks and calls for an in-depth reappraisal of conventional organizational design and management methods (Hamel, 2011; Ranson et al., 1980). This phenomenon is notably visible in how organizations use data and technology to seek a competitive advantage. Recent evidence from empirical data suggests that automation reduces routine manager tasks but redeliberates cognitive workload, as businesses with AI related technologies show enhanced productivity without being matched by commensurate manager monitoring (Domini et al., 2022). Speaking of AI-driven organizational form varies significantly over different market conditions, which correlate with different stages of technological adoption and institutional arrangements. Developing countries, by contrast, regularly construct innovative hybrids of their choice that help overcome some specific challenges relevant to them while avoiding the challenges pertinent to the existing state-of-the-art (Kumar & Puranam, 2012; Khanna & Palepu, 2010). As revealed by Government AI Readiness Index, developing countries show constant evidence as regards the above differences in trend because developed nations tend to keep sustaining high ranking indices concerning AI readiness (Oxford Insights, 2020-2024).

In what follows, this paper makes three interrelated theoretical contributions in depth through an organizational theory analysis of artificial intelligence capabilities and institutional environments. We explore how artificial intelligence helps organizations cut layers more effectively through superior coordination capabilities. We further elaborate on the 'hourglass model' as a novel theoretical concept for understanding novel organizational forms created by artificial intelligence, characterized by a broad strategic apex, a narrow middle tier, and a diversified operational base. We go on to explain how such structural changes vary systematically across developed and emerging market settings, depending on differences in technological readiness and in institutional environments.

Organizational-level trust in AI technologies is viewed as a prime influencer of AI-driven structural transformation, suggesting that considerations around transparency, dependability, and perceived autonomy are going to be important to the adoption of AI in management settings, for instance (Glikson and Woolley, 2020). Research studies on European firms have shown that AI-related innovations greatly affect gender-specific occupational segregation and human resource management policies (Belloc et al., 2022). These innovations have enabled the development of strong governance frameworks that guide AI-driven organizational change, including the European Data Sharing Agency for collaborative regulation of AI-based organizational systems (Graef & Prüfer, 2021).

This paper unfolds as follows: Section 2 describes the current state of AI capabilities and their implications for organizational structure. Our theoretical framework is elaborated upon in detail in Section 3, along with the mechanisms of AI-facilitated transformation. Section 4 covers the development of the concept of the hourglass structure, while Section 5 delineates hierarchical suppression and its market variations. Section 6 explores cross-market implementation patterns and institutional contexts. Conclusion discusses implications and future research directions.

## 2. Current AI Capabilities and Organizational Implications

AI systems' growing sophistication is transforming organizational information processing and decision-making, particularly for traditional managerial tasks (McAfee & Brynjolfsson, 2012). Advances in machine learning, natural language processing, and predictive analytics enable AI to perform diverse managerial tasks with unprecedented speed and accuracy (Davenport & Kirby, 2016; Grønsund & Aanestad, 2020). These include resource allocation and scheduling, monitoring, and performance assessment. It can monitor the performance of employees and give immediate feedback (Tambe et al., 2019). Empirical study results provide evidence for this point: compared with traditional management roles, in middle management, managers spend up to 54 percent of time on coordination and control activities that are mostly automatable (Kolbjørnsrud et al., 2016). The AI management systems have been implemented by the leading firms like IBM and Amazon for addressing issues of employee onboarding and productivity optimization assessment (IBM, 2018; Soper, 2021; Mirbabaie et al., 2021). It aligns with the traditional view of an organization as an information-processing system which may give rise to intelligent behavior (March, 1999; Ocasio, Rhee & Boynton, 2020).

The deployment landscape is heterogeneous across market-specific contexts. In developed economies, high labor costs and sophisticated digital infrastructures usually drive the rapid adoption of AI-based management systems (Brynjolfsson & McAfee, 2017). These organizations typically have mature data environments and established relationships with suppliers that facilitate AI technology adoption (Benbya et al., 2020). In developing economies, there are complex issues associated with adoption, as these countries must balance the requirement for technological advancement with infrastructural and maturity factors within their digital organizations

(Subramaniam et al., 2015). Lower labor costs in these markets alter the economic calculus for AI adoption and lead to more selective implementation in high-impact areas (Khanna & Palepu, 2010).

Recent studies have shown AI's growing sophistication in managing global value chains and optimizing resource allocation. These systems display three aspects that firms need to balance in their organizational transformation: autonomy in the automaticity of routine tasks, learning capacities for market intelligence, and combinative capabilities for resource reconfiguration (Hasan & Ojala, 2024). Generative AI has also been shown to be promising for strategic choice evaluation, particularly under conditions of complexity, which means it improves human choices when aggregated over several prompts and evaluation settings (Doshi et al., 2024).

However, the existing AI systems have some critical limitations that require serious attention. These systems are poor at decisions that demand subtle human factors and organizational culture, which are crucial ethical concerns related to accountability, fairness, and privacy (Jarrahi, 2018). The effective implementation of AI-based management systems requires a radical change management strategy to ensure successful human-machine integration and overcome resistance to change (Fountaine et al., 2019). These are more apparent in emerging markets, where hybrid designs are in demand to integrate AI with the local practices based on cultural and institutional contexts (Kumar & Puranam, 2012).

The future of AI is likely to expand its capabilities to handle increasingly complex managerial tasks because of technical advancement in reinforcement learning, causal inference, and AI ethics (Benbya et al., 2020). This progression indicates a path toward more advanced collaborative management systems involving both humans and artificial intelligence, in which AI assumes responsibility for routine and analytical functions while human managers concentrate on the strategic, creative, and interpersonal dimensions of leadership within organizations (Wilson & Daugherty, 2018; Singh & Anantatmula, 2013). Organizations that effectively integrate technological competencies with human considerations and the unique conditions of their markets will be optimally situated to leverage the potential advantages associated with AI-enhanced management automation (Mikalef & Gupta, 2021).

## 3. Theoretical Framework: Mechanisms of AI-Enabled Transformation

The integration of AI into organizational processes extends beyond technology adoption, fundamentally transforming organizational information processing capabilities and challenging core assumptions of traditional organizational theory (Puranam, 2018; Burton et al., 2020). Our analysis reveals three distinct mechanisms through which AI capabilities transform organizational functioning: algorithmic coordination, structural fluidity, and hybrid agency. Each mechanism is a product of the interaction between AI capabilities and organizational processes, giving rise to new possibilities for organizational design.

INSERT FIGURE 1 HERE

3.1 Algorithmic Coordination

Algorithmic coordination represents a paradigm shift from traditional organizational coordination methods. Unlike classical organizational theory's focus on hierarchical, interaction-based coordination (Thompson, 1967; Galbraith, 1974), it introduces a fundamentally new approach to ensuring organizational integration. Algorithmic coordination utilizes recent advances in artificial intelligence capabilities to achieve automated synchronization of the organization's functions through AI-driven prediction and optimization processes (Csaszar & Steinberger, 2021).

While human-mediated coordination faces cognitive and attention-based limitations, algorithmic coordination simultaneously processes multiple interdependencies across organizational boundaries. The very basis for traditional assumptions concerning organizational information processing and decision-making is changed hereby (March, 1999; Ocasio, Rhee & Boynton, 2020).

Empirical studies clearly indicate that algorithmic coordination greatly enhances organizational efficiency. It has been found that firms employing AI-related technologies experience a considerable increase in productivity without an increase in managerial control (Domini et al., 2022). The underlying cause for such increased efficiency is that AI enables more simultaneous handling of many interdependencies, making processes more efficient and either maintaining or improving the quality of coordination (Kolbjørnsrud, 2024). These findings challenge existing assumptions about the relationship between managerial control and organizational performance.

Thus we propose:

P1: Organizations adopting AI-enabled algorithmic coordination demonstrate increased operational efficiency without proportional increases in managerial oversight.

3.2 Structural Fluidity

Structural fluidity emerges as the second key mechanism through which AI drives organizational transformation. We define structural fluidity as an organization's capacity to dynamically reconfigure its structure in response to external environmental pressures. This mechanism goes beyond the traditional organizational flexibility concepts since it incorporates AI's unique capability of allowing instantaneous adjustments in structure (Reeves et al., 2015, 2018).

AI's integrative capabilities allow managers to reshape value-generating activities and develop unique dynamic capabilities, creating firm-specific advantages (Hasan & Ojala, 2024). This effect is particularly dominant in developed markets characterized by higher levels of digital maturity and a greater propensity towards flat hierarchies (Bernstein et al., 2016). By contrast, in emerging markets, flexibility often manifests at the structural level by way of hybrid organizational forms that reflect different institutional and cultural backgrounds (Khanna & Palepu, 2010).

Structural fluidity's dynamic reconfiguration capabilities create new opportunities for organizational adaptation while maintaining operational stability. AI-facilitated coordination mechanisms enable organizations to respond rapidly to environmental changes while preserving core functions.

Thus we propose:

P2: Organizations exhibiting higher structural fluidity will demonstrate superior adaptability to environmental changes while maintaining operational stability through AI-enabled coordination mechanisms.

3.3 Hybrid Agency

Hybrid agency is the third mechanism and forms a new theoretical concept that captures the distinctive nature of decision-making in AI-augmented organizations. Beyond the human-machine interaction model, hybrid agency involves a more complex interplay between human strategic cognition and AI analytical capabilities. A recent study demonstrated how generative AI changes the game in organizational knowledge management, enabling synthesis and creative abilities but involving some risks like AI-generated inaccuracies (Alavi et al., 2024).

Hybrid agency is very important in organizational functioning and is contingent on the degree of organizational trust in AI technologies, such that attributes like transparency, reliability, and perceived autonomy are essential preconditions for having AI implemented in management situations (Glikson & Woolley, 2020; Kaplan et al., 2023). This conceptual framework details how organizations can leverage the strengths of people and AI systems to stay fit strategically.

Thus we propose:

P3: The effectiveness of hybrid agency in AI-augmented organizations is positively related to organizational trust in AI technologies.

These three mechanisms, namely, algorithmic coordination, structural fluidity, and hybrid agency, form a theoretical basis to understand the transformative influence of AI on the organizational structures. Through this theoretical construct, the existing organizational theories are broadened to explain how AI promotes new forms of organizations that go beyond organizational structural restrictions. The expression of these mechanisms turns out to be different in various market environments, indicating varying levels of technological preparedness and institutional support (Kumar & Puranam, 2012).

This understanding of theoretical mechanisms and their relationships constitutes the groundwork for grasping how hourglass structures can appear as well as how hierarchical suppression emerges within contemporary organizations. The conceptualization of the mechanism reveals the sources for the potential divergences of success for different organizations working under varying market environments for AI-driven change, thus offering insight to manage its successful introduction across the array of institutional contexts.

INSERT FIGURE 2 HERE

## 4. The Hourglass Structure: A New Organizational Paradigm

The increasing adoption of AI in traditional middle management functions drives a fundamental transformation from pyramid to 'hourglass' structures. This shift represents not merely incremental change but a foundational reconceptualization of organizational design in an AI-augmented world. This is more than incremental change; it reflects a foundational recasting of organizational design in an AI-augmented world (Aghina et al., 2018; Boos, 2024). A classic information flow facilitator and a decision-making tool for big organizations-the small apex of senior executives, the broad middle layer of managers, and the wide bottom layer of front-line workers-constitutes a traditional pyramid structure (Mintzberg, 1979; Robertson, 2015). However, it is increasingly at odds with the nature and needs of AI-driven operations, as described in the following text.

The AI-driven hourglass structure comprises three distinct layers. At the top, expanded senior leadership focuses on strategic roles, enabled by AI's handling of operational tasks. The middle section is compressed, with fewer human middle managers but enhanced by AI systems that handle traditional coordinating functions. The base remains broad but becomes more diverse, encompassing front-line workers, specialists, and integrated AI systems (Shrestha et al., 2019). This structural change is in line with the current dynamics of labor markets, as AI disproportionately affects middle-skill jobs while complementing both high-skill and low-skill positions (Tschang & Almirall, 2021).

INSERT FIGURE 3 HERE

New empirical evidences show how AI shifts cognitive responsibilities within organizations. Firms using AI-associated technologies record higher productivity growth. However, its associated managerial monitoring does not necessarily increase; the firms instead focuses on maximizing the operations and productivity (Domini et al., 2022). The effects of AI complementarity are now arising in labor markets. The advanced economies have already recorded more jobs in AI-driven careers (Acemoglu & Restrepo, 2018). While AI integration is likely to be most valuable for cognitive-intensive jobs, displacement risks are particularly high for non-college-educated and older workers (IMF Staff, 2024).

The transition to an hourglass structure is very different across market contexts. In developed markets, which have higher digital maturity and a greater acceptance of flat hierarchies (Bernstein et al., 2016), organizations are more likely to adopt the pure hourglass model. New studies indicate that the combinatory properties of artificial intelligence enable managers to reshape value-creating activities while developing dynamic capabilities leading to firm-specific advantages (Hasan & Ojala, 2024). In contrast, emerging markets are more likely to present hybrid structures characteristic of their institutional and cultural settings (Khanna & Palepu, 2010). These hybrid configurations tend to maintain some traditional hierarchical characteristics but deliberately integrate AI-driven processes, especially in cultures where interpersonal dynamics and hierarchical reverence are still paramount (Xin & Pearce, 1996).

Thus we propose:

P4: Market context moderates hourglass structure implementation, with developed markets showing more complete adoption while emerging markets demonstrate hybrid variations that reflect local institutional contexts.

Core competencies of multinational corporations that are relevant to the AI age include the technological and cognitive aspects of AI (Jaiswal et al., 2021). Basic technological competencies include data analysis capabilities and digital literacy, while basic cognitive skills include sophisticated processing, decision-making capabilities, and a lifelong learning attitude. Even in developing countries like India, which have high rates of AI adoption, organizations are still fostering hybrid models that balance the advancement of technology with the traits of the labor force and local cultural values (Chatterjee, 2020).

Structural change has critical implications for career development and organizational behavior. The evolution of career development frameworks within organizations enhanced by artificial intelligence introduces distinct challenges and prospects. Organizations are required to formulate novel advancement pathways that acknowledge both technical proficiencies and leadership competencies, while also considering the diminished availability of conventional management roles (Chamorro-Premuzic et al., 2018). It is essential for research to investigate how organizations can establish significant career paths that correspond with individual ambitions and organizational requirements in AI-augmented environments (Wilson & Daugherty, 2018).

Although the hourglass framework promotes greater efficiency and flexibility, its effective application necessitates considerable investment in technology as well as in change management practices (Tarafdar et al., 2019). With the ongoing advancements in artificial intelligence capabilities, it is probable that organizational structures will evolve to become more dynamic and attuned to environmental requirements (Bernstein et al., 2016). Therefore, the hourglass model should be viewed not as a terminal stage but as part of an ongoing cycle of change within organizational structure under the relentless impulse of technological improvements and shifts in regional market circumstances.

## 5. Hierarchical Suppression and Market Variations

AI enables "hierarchical suppression"—a systematic reduction in organizational levels that fundamentally transforms traditional management structures. This is an extension and acceleration of trends toward flatter organizations, where research shows that organizations and AI systems have to solve many similar coordination problems, which makes AI especially well suited to assuming traditional hierarchical coordination functions (Csaszar & Steinberger, 2021).

Through sophisticated information processing capabilities, AI networks transform organizational communication patterns. By efficiently collecting, processing, and sharing information across the organization, these systems reduce the need for multiple layers of human information processing. This capability undermines one of the primary historical justifications for hierarchical organizations: the need to filter and transmit information up and down the organizational chain. Artificial intelligence can provide real-time and extensive data analytics to decision-makers at any level of an organization, rendering the traditional function of middle managers as information intermediaries progressively redundant (Davenport, 2018).

Hierarchical compression through AI typically leads to several unique characteristics. Organizations usually have more flat, project-based team structures instead of hierarchical divisions between departments (Aghina et al., 2018). Meanwhile, decision-making processes are now being decentralized and data-driven; with artificial intelligence, decisions at every level are now made using real-time analytics (Davenport, 2018). The flattening of organizational hierarchies through artificial intelligence has several advantages: more agile organizations can respond more quickly to changes in the market, and direct communication between top management and front-line employees can help to create more integrated organizations, and information distortion is also decreased (Hamel, 2011; Rajan & Wulf, 2006).

The transformation of hierarchical structures has to be grasped in varying market contexts. As Janssen et al. (2020) observe, in developed markets, high middle management costs and advanced technological infrastructure may accelerate this process. This raises important considerations about control and responsibility. Hasan and Ojala (2024) note that the autonomy of AI systems creates challenges in attributing responsibility and liability, particularly in international operations where institutional frameworks for AI governance vary significantly across countries.

In emerging markets, traditional hierarchical relationships often play critical substitute roles for formal institutional support (Xin & Pearce, 1996), suggesting that hierarchical suppression might be a spectrum rather than a binary state (Puranam et al., 2014).

Thus we propose:

P5: The degree of hierarchical suppression will vary systematically across market contexts, with organizations in developed markets exhibiting higher levels of AI-enabled structural flattening compared to emerging markets, where hierarchical relationships serve crucial institutional functions.

Organizations differ in this aspect as based on the market and the related institutional context (Wadhwa et al., 2008; Rao & Sutton, 2014). These processes start progressing at an appreciably more incremental rate as lower labor costs, cultural preferences that support pyramid structures, and varying levels of preparation to accept new technologies influence the same emerging markets (Sharma & Good, 2013).

However, hierarchical suppression also has major risks and challenges that need to be approached with caution. Over-flattening may result in too little human oversight in key areas (Shrestha et al., 2019), and the transition may disrupt organizational culture, which is especially damaging in companies with a history of hierarchy (Fountaine et al., 2019). Regular career advancement will not be readily available, thereby affecting employee motivation and retention at work (Chamorro-Premuzic et al., 2018). While AI can perform many coordination tasks, complex, nuanced coordination may still require human intervention, which can be difficult in very flat structures (Jarrahi, 2018).

The existence of hierarchical suppression suggests a spectrum that is not binary but rather complex, with differences across different organizational environments and their capacities (Puranam et al., 2014). This spectrum includes three interrelated dimensions of transformation. First, the level of AI integration ranges from simply augmenting existing processes to fundamentally reconfiguring organizational coordination structures (Shrestha et al., 2019). Thus, organizations that are on the lower end of this range use AI tools to improve their current hierarchical processes, while those on the higher ends make use of AI systems to fundamentally redesign their coordination methods (Davenport, 2018; Krakowski et al., 2022).

The patterns of structural adaptation also reflect varying degrees of hierarchical compression. Some organizations maintain traditional layering with minimal compression, while others adopt hybrid structures that deliberately compress management layers to match functional requirements and AI capabilities (Bernstein et al., 2016). The most advanced applications fully adopt an hourglass structure, characterized by strong compression of the middle layer and enhanced coordination through AI systems (Fountaine et al., 2019; Wilson & Daugherty, 2018).

Third, market context significantly influences the progression of hierarchical suppression. Organizations in developed markets typically advance more rapidly along this spectrum, driven by higher labor costs and more mature technological infrastructure (Brynjolfsson & McAfee, 2017). In contrast, emerging market organizations often demonstrate more gradual and selective adaptation, influenced by institutional factors and cultural preferences for hierarchical relationships (Kumar & Puranam, 2012; Xin & Pearce, 1996). This variation in implementation speed and depth reflects broader patterns of technological adoption and organizational change across different market contexts (Luo, 2021).

Successful implementation of hierarchical suppression involves planning and change management, including wide-ranging reskilling initiatives (World Economic Forum, 2020), clear roles and decision-making authorities definition (Bernstein et al., 2016), and active management of cultural transitions (Fountaine et al., 2019; Vincent, 2021). As AI capabilities evolve, we are likely to encounter even more drastic forms of hierarchical compression, with organizational structures verging on the "liquid" according to some futurists (Johansen, 2017). Nevertheless, the flattening will be taken too far for some types of organizations. The challenge to organizational leaders and researchers alike will be in striking the right balance of human and AI-driven management within flatter structures (Wilson & Daugherty, 2018) while ensuring the benefits of hierarchical suppression are achieved at no cost to organizational effectiveness and human well-being.

## 6. Cross-Market Implementation and Institutional Contexts

The manifestation of AI-enabled organizational transformation varies systematically across different market contexts, reflecting deep institutional and structural differences between developed and emerging markets. This variation extends beyond simple differences in adoption rates or implementation capabilities, reflecting deeper relationships between institutional arrangements, technological capabilities, and organizational transformation processes (Luo, 2021).

INSERT FIGURE 4 HERE

Developed economies, characterized by high labor costs and sophisticated digital infrastructures, typically demonstrate rapid adoption of AI-based management solutions (Brynjolfsson & McAfee, 2017). These organizations leverage mature data ecosystems and established vendor relationships to facilitate AI integration (Benbya et al., 2020). Technological advancements have enabled this shift from traditional enterprise computing, which is efficiency-oriented, to a modern pervasively computing world that integrates virtual and physical workspaces seamlessly (Cascio & Montealegre, 2016).

Emerging markets face unique implementation issues that demand context-specific solutions. Some technology hotspots, like in India, show high levels of AI maturity; however, other aspects of organizational change are typically limited by the infrastructure and level of digital maturity in these markets (Kumar & Puranam, 2012). Most organizations in emerging markets adopt some form of hybrid solution, where AI-based processes are selectively implemented in high-impact areas, and traditional structures remain in place wherever culturally acceptable (Khanna & Palepu, 2010; Cappelli et al., 2010).

Cultural factors significantly shape organizational change across market environments. Advanced markets typically embrace decentralized systems and autonomous decision-making, while developing markets tend to maintain traditional management practices and hierarchical structures (George et al., 2014). This difference is also statistically expressed in the Government AI Readiness Index (2020-2024), where it points to long-standing artificial intelligence readiness gaps between developed and emerging markets.

Studies with companies in Europe indicated that AI-related innovations exert effects that are of significant magnitude on gender-based job segregation and human resource management policies (Belloc et al., 2022). Against this background of impacts on the labor market, there is an increase in robust frameworks of governance serving as a necessary driver for steering AI-driven change in organizational direction. Major activities entail the establishment of a body called the European Data Sharing Agency, which enables collaborative monitoring of AI-based organizational systems (Graef & Prüfer, 2021).

The structure of institutions significantly determines the way in which organizations incorporate and deploy AI capabilities (Janssen et al., 2020). Organizations respond to institutional pressures through three crucial mechanisms: coercive pressures originating from regulatory compliance and industry expectations, mimetic pressures to mimic successful AI implementations as seen in peer organizations, and normative pressures from professional expectations of how AI should be used (DiMaggio & Powell, 1983; Caplan & Boyd, 2018).

Thus we propose:

P6: Institutional pressures shape organizational AI adoption through three mechanisms: coercive pressures from regulations, mimetic pressures from peer organizations, and normative pressures from professional expectations.

Global activities related to artificial intelligence have unique challenges, which are related to digital interconnectivity and cybersecurity. These risks demand sophisticated information-processing capabilities in order to cope with uncertainty and vulnerabilities effectively (Luo, 2021). The architecture of regulations surrounding AI deployment is highly heterogeneous since developed markets tend to operate under more mature regimes of AI governance and data protection (Graef & Prüfer, 2021). Although emerging markets often embrace more flexible regulatory approaches, they must eventually develop robust governance systems, which will influence the pace and nature of organizational transformation (Manyika et al., 2017).

Recent empirical studies have revealed the development of increasingly sophisticated artificial intelligence capabilities in different managerial domains. These capabilities express themselves through three clear but interconnected dimensions, namely, autonomy in routine task automation, learning capabilities for market understanding, and combinative features for resource reconfiguration (Hasan & Ojala, 2024). This level of effectiveness, however, relies on the organizations' trust of AI-based technology; in this regard, such aspects as transparency, dependability, and perceived autonomy are the most critical aspects that determine whether AI is included in management environments (Glikson & Woolley, 2020; Kaplan et al., 2023).

Thus we propose:

P7: The effectiveness of AI implementation is determined by three key organizational capabilities: autonomy in routine task automation, learning capabilities for market understanding, and combinative features for resource reconfiguration.

Effective cross-market implementation requires a holistic approach that involves technological, human, ethical, and social aspects. Organizations must simultaneously manage workforce transitions, leadership challenges, and ethical issues while maintaining operational efficiency and organizational cohesion. Their ability to navigate these challenges and respond to market-specific conditions will determine their success in the age of artificial intelligence (Wilson & Daugherty, 2018).

INSERT FIGURE 5 HERE

## 7. Discussion and Implications

The organizational integration of AI presents implications that extend far beyond conventional technology adoption concerns. For example, in the operation of it, embedding AI greatly hastens the pace while at the same time maximizes the efficacy of decision-making (Krakowski et al., 2022), allowing organizations to process large quantities of data at a velocity significantly higher than human-centric approaches alone (Brynjolfsson & McAfee, 2017). This advanced capability helps firms better forecast and respond to shifts in market demand ahead of the competition through predictive analytics; however, it is imperative that appropriate controls and monitoring measures be established for organizations so that the enhanced speed does not compromise the integrity of decisions or standards (Agrawal et al., 2018; Shrestha et al., 2019).

The AI-driven hourglass framework introduces new dynamics in organizational agility and cost management. Agile teams, which are established for immediate competencies and requirements, allow organizations to reconfigure more efficiently in response to market fluctuations (Bernstein et al., 2016). While such changes would also bring about savings in overhead costs significantly, the organization will have to weigh these savings against the very high costs of the required investments in artificial intelligence technologies and the management of workforce transitions (Fountaine et al., 2019).

The workforce implications require particular attention and strategic planning. Organizations must develop comprehensive reskilling initiatives to prepare employees for AI-augmented environments, focusing on critical competencies such as data literacy, AI interaction, and complex problem-solving (World Economic Forum, 2020). Although restructuring due to AI may eventually displace many middle managers, the development of AI itself produces new jobs related to the management of AI systems, human-AI interaction, and AI ethics (Davenport & Kirby, 2016; Wilson & Daugherty, 2018; Frey & Osborne, 2017). Among other things, there are deep social and economic implications, including increasing inequality, besides these new social safety nets (Korinek & Stiglitz, 2017).

Successfully managing AI-integrated, flatter organizations requires transformed leadership roles. Leaders must develop AI literacy while focusing more intensively on long-term strategy and vision-setting. Effective oversight of human-AI collaboration demands balancing critical thinking with AI system trust, while preserving human agency and creativity (Jarrahi, 2018). For instance, generative AI transforms organizational knowledge management by improving synthesis and creative capabilities but underscores specific risks such as inaccuracies through AI-generated information while pointing to the need for strong governance frameworks (Alavi et al., 2024).

Ethical concerns become more relevant as AI systems begin to take over more managerial tasks. Organizations have to deal with the issue of bias in AI decision-making, ensure transparency and accountability, and protect privacy while meeting substantial data requirements (Tambe et al., 2019). This calls for the development of robust frameworks for ethical AI use and regular system audits, navigating complex data protection regulations across jurisdictions (Floridi et al., 2018).

The extent of AI-led business transformation differs sharply between developed and emerging economies. In the developed world, mature markets are confronted with high labor costs and already built-up digital architecture that strongly encourages the deep penetration of AI. These factors are primarily supported by long-standing relationships with vendors and advanced data infrastructures (Brynjolfsson & McAfee, 2017). For example, US firms have embraced AI-based management applications quickly in sectors where the costs of middle management are high (Tambe et al., 2019).

Emerging markets face distinct implementation issues, necessitating great caution while taking into consideration local conditions. Even though some technology hubs, such as those in India, exhibit leading AI capabilities, other organizational change generally faces infrastructure and digital maturity differences (Kumar & Puranam, 2012). In most emerging markets, organizations have invented unique hybrid approaches: selectively adopting AI-based processes for the high impact areas while leaving the rest as it is with traditional structures because that's what still works for them (Khanna & Palepu, 2010).

Cultural factors are key drivers of organizational change in any given market environment. Developed markets tend to have a greater receptiveness to flattened organizational structures and self-directed decision-making processes. In contrast, emerging markets tend to favor conventional management practices and hierarchical relationships with others (George et al., 2014). This is further quantitatively reflected in the Government AI Readiness Index for the period between 2020 and 2024, indicating that developed and emerging markets still have different AI preparedness.

## 8. Future Research Directions

The hourglass model of AI-augmented organizations presents critical research opportunities for systematic investigation. The transition from traditional to hourglass structures raises fundamental questions about organizational design and adaptation mechanisms. Therefore, research is required to explore how organizations adapt to this structural change, especially by considering the interplay between technological capabilities, organizational readiness, and implementation strategies (Benbya et al., 2020; Krakowski et al., 2022).

A key research priority is examining the performance implications of hourglass structure adoption. Future studies should investigate how this structural transformation affects various organizational metrics, including operational efficiency, innovation capacity, and adaptive capabilities (Mikalef & Gupta, 2021). The human capital implications are particularly relevant, especially in terms of how organizations maintain employee engagement and development in flattened structures (Chamorro-Premuzic et al., 2018; Tambe et al., 2019).

Another crucial research area involves examining implementation patterns across organizational contexts. Researchers should investigate how organizations adapt the hourglass model to meet specific industry needs and market requirements (Janssen et al., 2020). Hybrid forms in various market environments have increased demand for careful studies of regulatory settings and institutional contexts that influence the structural choice (Graef & Prüfer, 2021). Understanding the above diversity will lead to critical findings that can be of value for both theory development and practice.

Hybrid organizational frameworks in developing economies deserve substantial attention due to their rise and effectiveness. In most of these areas, there are innovative approaches toward aligning technological advancements with employment trends and cultural forces (Khanna & Palepu, 2010). There is evidence that emerging markets develop unique solutions that pay homage to the traditional hierarchical structures but increasingly incorporate AI-facilitated processes in areas where they can provide maximum benefits (Xin & Pearce, 1996). Then, future studies should analyze how these hybrid models evolve and their outcomes on organizational performance.

The other important area of research is the relationship between labor costs, technological readiness, and AI adoption rates. Studies involving European companies have shown that AI-related innovations do have significant effects on gender-based job segregation and HRM policies (Belloc et al., 2022). Furthermore, evidence has shown that job functions that require high levels of cognitive intensity would be enhanced by AI; however, the risk of displacement is still relatively high for non-college-educated workers and the elderly workers (IMF Staff, 2024). Future studies should look into how such patterns vary within market contexts and what implications these carry to workforce development.

Leadership competencies in AI-augmented organizations require systematic examination. New competencies are needed for leaders in AI-integrated, flatter organizations with strategic focus and human creativity (Jarrahi, 2018). Recent research identified key skills for the AI era, especially in technological and cognitive domains (Jaiswal et al., 2021; Trunk et al., 2020). Future research could investigate how development programs for leaders could effectively develop such competencies in culturally cohesive ways while operating in more AI-driven environments.

It has become a high research priority to establish ethical frameworks for the governance of artificial intelligence (Xue & Pang, 2022). Organizations must deal with bias in AI decision-making, provide transparency and accountability, and safeguard privacy while ensuring the huge amount of data that AI systems require (Tambe et al., 2019). Governance structures such as the European Data Sharing Agency develop foundational models for cooperative oversight of AI-integrated organizational systems (Graef & Prüfer, 2021). Future studies should evaluate the effectiveness of these models and their applicability in different institutional contexts.

Mechanisms of knowledge transfer and best practices for AI-driven transformation between developed and emerging markets need to be researched. Though technology hubs may boast cutting-edge AI, overall organizational transformation suffers from the lack of infrastructure and digital maturity in many organizations (Kumar & Puranam, 2012). Knowing how to effectively transfer and adapt knowledge about AI implementation between market contexts remains an important area of future research.

Further research is required on the development of career paths in hourglass structures. The traditional career ladders, which rely on climbing the different management levels, need to be overhauled completely to address the bottlenecked middle layer effectively (Chamorro-Premuzic et al., 2018). Future studies should be aimed at how organizations can develop workable alternative career progression pathways that maintain employee engagement and retention in AI-enhanced environments.

Another important research direction is the role of artificial intelligence in organizational learning and knowledge development. Recent studies have shown how generative AI transforms organizational knowledge management by improving synthesis and creative capabilities while introducing specific risks through AI-generated information (Alavi et al., 2024). Future research should look into how organizations can effectively leverage AI for knowledge creation while managing associated risks and maintaining the quality of organizational learning.

**Statements and Declarations**

**Author Contributions:** Both authors, K.K. Balaraman and V.R.R. Ganuthula, contributed equally at all stages of research leading to manuscript submission.

**Funding Statement:** No specific grants received.

**Conflict of Interest:** No conflicts declared.

**Data Availability:** No new data reported.

**Software and AI Usage Statement:** The authors used Claude Sonnet 4.6 (www.claude.ai) to assist with structuring the manuscript and shaping its narrative flow. Grammarly (www.grammarly.com) was used for grammar and style improvements. The authors reviewed and edited all content and take full responsibility for the final article.

35. IMF Staff. (2024). Employment, technology, and growth in the 21st century (Working Paper No. WP/24/15). International Monetary Fund.

36. Jaiswal, A., Arun, C. J., & Varma, A. (2021). Rethinking managerial capabilities in the age of AI: Evidence from Indian firms. *Journal of Business Research, 128*, 704-718.

37. Janssen, M., Weerakkody, V., Ismagilova, E., Sivarajah, U., & Irani, Z. (2020). A framework for analysing blockchain technology adoption: Integrating institutional, market and technical factors. *International journal of Information Management*, *50*, 302-309.

38. Jarrahi, M. H. (2018). Artificial intelligence and the future of work: Human-AI symbiosis in organizational decision making. *Business Horizons, 61*(4), 577-586.

39. Johansen, R. (2017). *The new leadership literacies: Thriving in a future of extreme disruption and distributed everything*. Berrett-Koehler Publishers.

40. Kaplan, A. D., Kessler, T. T., Brill, J. C., & Hancock, P. A. (2023). Trust in artificial intelligence: Meta-analytic findings. *Human factors*, *65*(2), 337-359.

41. Khanna, T., & Palepu, K. G. (2010). *Winning in emerging markets: A road map for strategy and execution*. Harvard Business Press.

42. Kolbjørnsrud, V. (2024). Designing the Intelligent Organization: six principles for Human-AI collaboration. *California Management Review*, *66*(2), 44-64.

43. Kolbjørnsrud, V., Amico, R., & Thomas, R. J. (2016). How artificial intelligence will redefine management. *Harvard Business Review Digital Articles*, 2-11.

44. Korinek, A., & Stiglitz, J. E. (2017). Artificial intelligence and its implications for income distribution and unemployment. *NBER Working Paper No. 24174*.

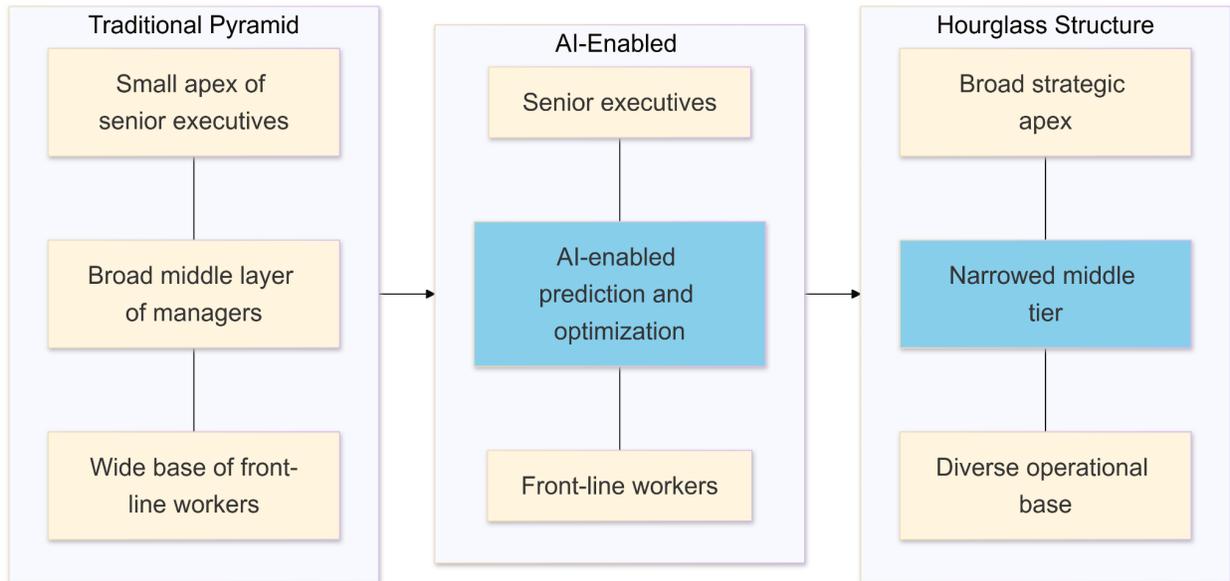

**Figure 1:** Evolution of Organizational Forms in the AI Era

Figure 1 illustrates the evolution from traditional hierarchical structures to the AI-enabled hourglass form, highlighting the key structural transformations.

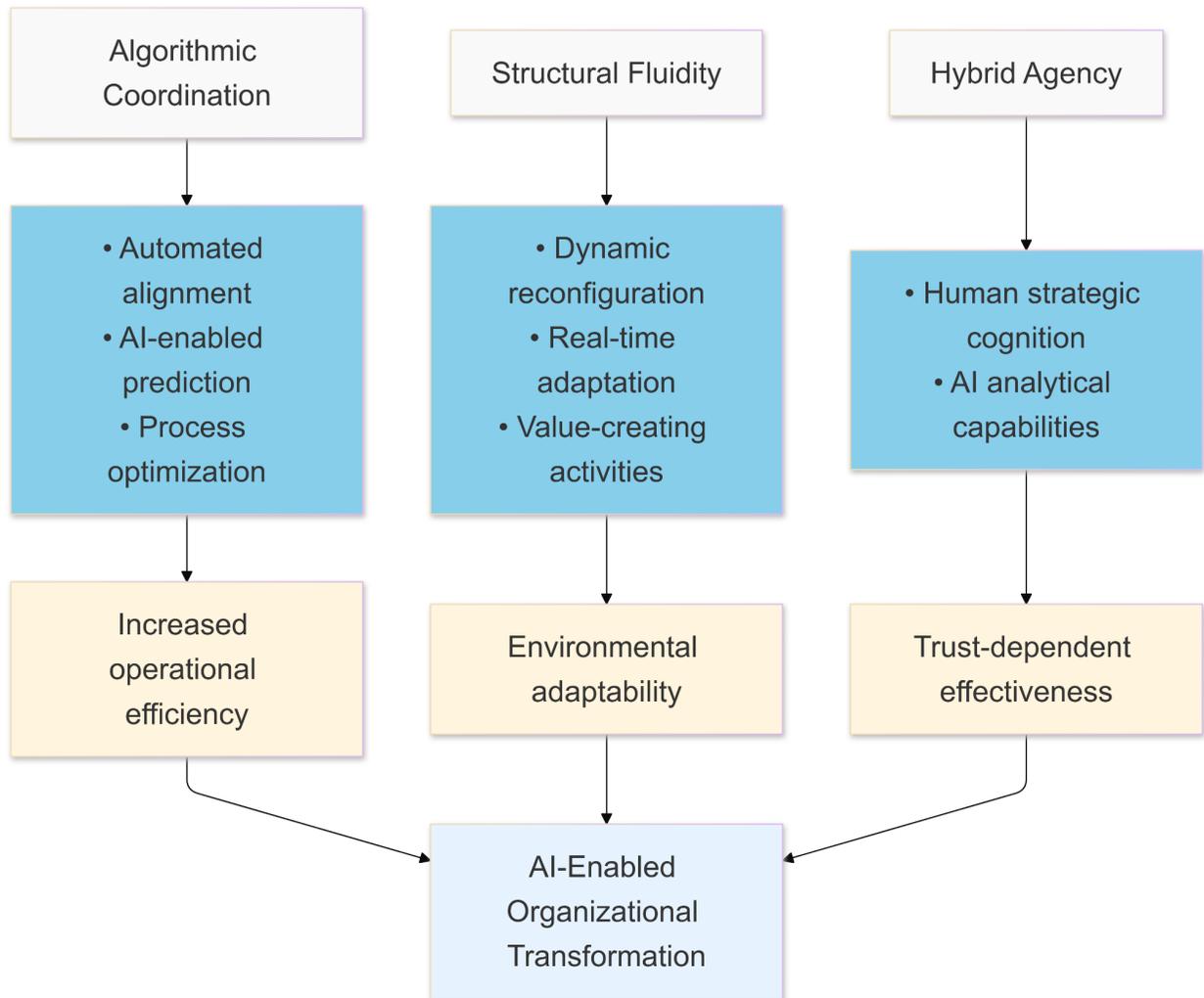

**Figure 2:** Mechanisms of AI-Enabled Organizational Transformation

Figure 2 illustrates the three core mechanisms of AI-enabled transformation—algorithmic coordination, structural fluidity, and hybrid agency—showing their key features and organizational outcomes.

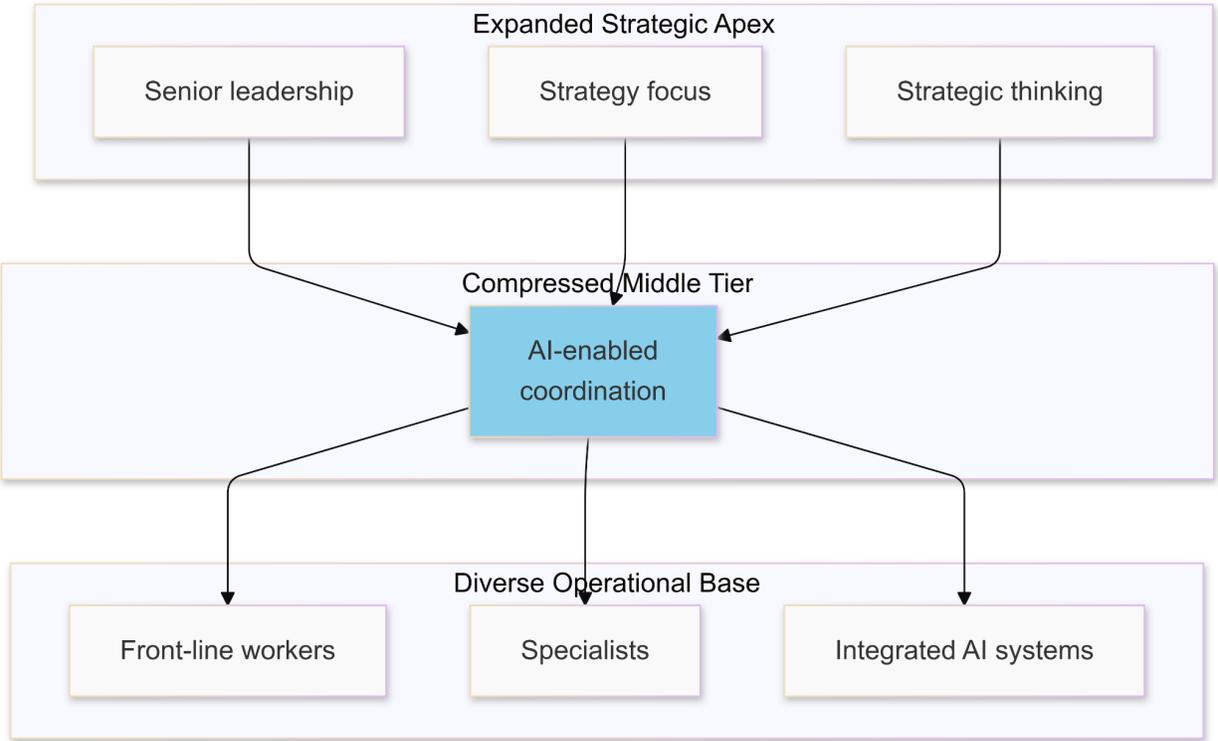

**Figure 3:** The Hourglass Structure Components

Figure 3 details the components and functions of each layer in the hourglass structure.

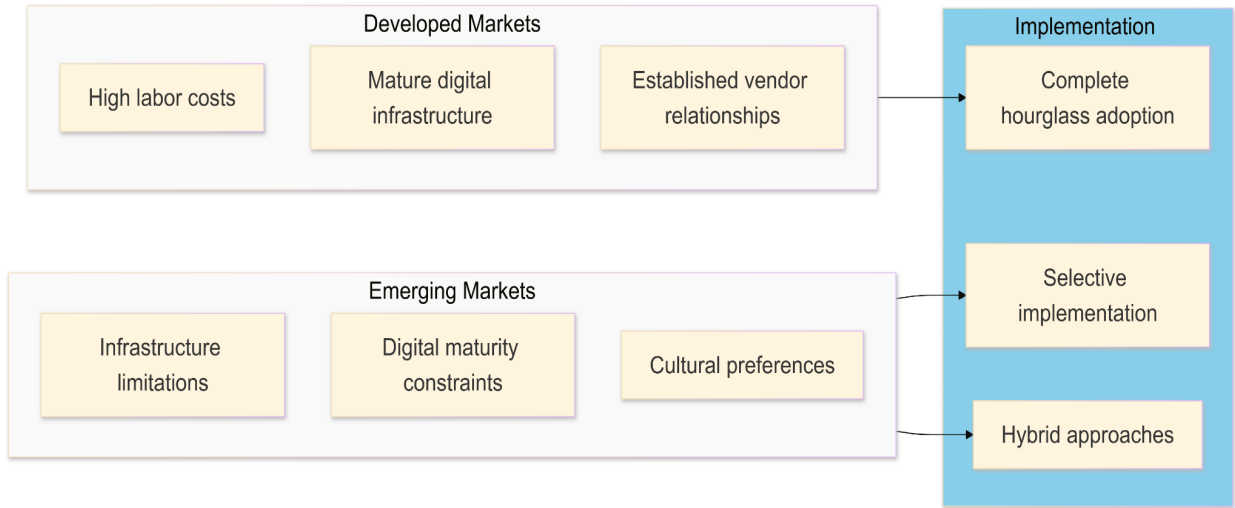

**Figure 4:** Market Context Implementation Patterns

Figure 4 contrasts the implementation patterns and characteristics across developed and emerging markets.

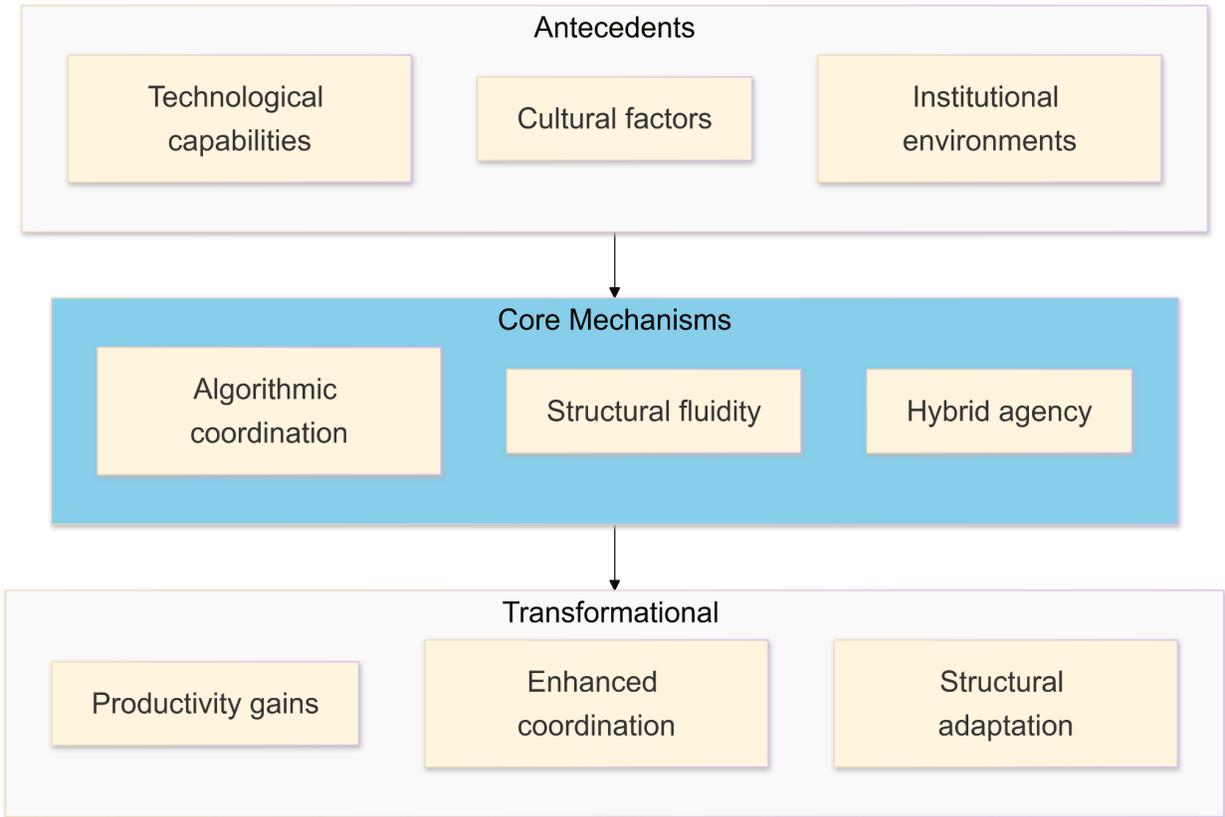

**Figure 5:** Theoretical Framework of AI-Enabled Organizational Transformation

Figure 5 synthesizes the theoretical framework, showing the relationships between antecedents, mechanisms, and outcomes of AI-enabled organizational transformation.